\documentclass[aps,prl,preprintnumbers,showpacs,twocolumn,nofootinbib]{revtex4}
\usepackage{graphicx}
\usepackage{dcolumn}
\usepackage{bm}
\newcommand{\beq}{\begin{equation}}
\newcommand{\eeq}{\end{equation}}
\newcommand{\beqa}{\begin{eqnarray}}
\newcommand{\eeqa}{\end{eqnarray}}

\newcommand{\pla}{\text{Pl}}

\def\beq{\begin{equation}}
\def\eeq{\end{equation}}
\def\bea{\begin{eqnarray}}
\def\eea{\end{eqnarray}}
\newcommand{\be}{\begin{equation}}

\newcommand{\ee}{\end{equation}}
\def\beqa{\begin{eqnarray}}
\def\eeqa{\end{eqnarray}}
\def\beq{\begin{equation}}
\def\eeq{\end{equation}}

\def\gd{g_{\mu\nu}}

\def\dab{_{\alpha\beta}}

\let\gam=w

\def\gd{g_{\mu\nu}}

\renewcommand{\epsilon}{\varepsilon}

\def\I{{\cal I}}
\newcommand{\tev}{\text{TeV}}
\newcommand{\gev}{\text{GeV}}
\newcommand{\ev}{\text{eV}}
\newcommand{\kev}{\text{keV}}

\newcommand{\Mev}{\text{MeV}}
\newcommand{\s}{\text{s}}
\newcommand{\cm}{\text{cm}}
\newcommand{\bbn}{\text{BBN}}

\begin{document}

\preprint{FTPI-MINN-08/35}

\preprint{UMN-TH-2716/08}

\preprint{arXiv:0809.1653}

\title{Dark Matter from $R^2$-gravity}

\author{Jose A. R. Cembranos}

\affiliation{
William I. Fine Theoretical Physics Institute,
University of Minnesota, Minneapolis, 55455, USA
}

\begin{abstract}
The modification of Einstein gravity at high energies is mandatory from
a quantum approach. In this work, we point out that this modification
will necessarily introduce new degrees of freedom. We analyze the possibility
that these new gravitational states can provide the main contribution to the 
non-baryonic dark matter of the Universe. Unfortunately, the right 
ultraviolet completion of gravity is still unresolved. For this reason, we will
illustrate this idea with the simplest high energy modification of the Einstein-Hilbert 
action: $R^2$-gravity.

\end{abstract}


\maketitle

\section{Introduction}

Different astrophysical observations agree that the main amount of the matter content of our Universe is in form of unknown particles that are not included in the Standard Model (SM). Typical candidates to account for the missing matter can be found in well motivated extensions of the electro-weak sector. However, there is a fundamental sector in our model of particles and interactions, where the introduction of new degrees of freedom is not only well motivated, but absolutely necessary. 

The non-unitarity and non-renormalizability of the gravitational interaction described by the Einstein-Hilbert action (EHA) demands its modification at
high energies. The main idea of this work is to realize that this correction cannot be accomplished without the introduction of new states; these states will typically interact with SM fields through Planck scale suppressed couplings and potentially work as dark matter (DM).

\section{$R^2$-gravity}

In spite of many and continuous efforts, the ultraviolet (UV) completion of the gravitational interaction is still an open question. In these conditions, it is difficult to make general statements about its phenomenology. We will adopt a very conservative and minimal approach in order to capture the fundamental physics of the problem. The simplest correction to the EHA at high energies is provided by the inclusion of four-derivative terms in the metric that preserve the general covariance principle. The most general four-derivative action supports, in addition to the usual massless spin-two graviton, a massive spin-two and a massive scalar mode, with a total of eight degrees of freedom (in the {\it physical} or {\it transverse} gauge \cite{Stelle:1976gc, Stelle:1977ry}). In fact, four-derivative gravity is renormalizable. However, the massive spin-two gravitons are ghost-like particles that generate new unitarity violations, breaking of causality, and inadmissible instabilities \cite{Simon:1991bm}. 

In any case, in four dimensions, there is a non-trivial four-derivative extension of Einstein gravity that is free of ghosts and phenomenologically viable. It is the so called $R^2$-gravity since it is defined by the only addition of a term proportional to the square of the scalar curvature to the EHA.  This term does not improve the UV problems of Einstein gravity but illustrates our idea in a minimal way.
In fact, $R^2$-gravity only introduces one additional scalar degree of freedom, whose mass $m_0$ is given by the corresponding new constant in the action, as one can see in Eq. (\ref{model}):
\begin{eqnarray}
\label{model}        
S_G&=&\int\sqrt{g}\left\{-\Lambda^4
-\frac{M_{\pla}^2}{2}R
\,+\frac{M_{\pla}^2}{12\,m_0^2}R^2
\,+\,...\,\right\}
\\
&&
\;\;\;\;\;
\;\;\;\;\;\;\;\;\underbrace{\;\;}_{\text{DE}}
\;\;\;\underbrace{\;\;\;\;\;\;\;\;\;\;\,}_{\text{EHA}}
\;\;\;\;\underbrace{\;\;\;\;\;\;\;\;\;\;\;\;\;\;}_{\text{DM}}\nonumber
\end{eqnarray}
where $M_{\pla}\equiv (8\pi G_N)^{-1/2}\simeq 2.4 \times 10^{18}$ GeV,
$\Lambda\simeq 2.3\times 10^{-3}$ eV, and the dots refer to higher energy corrections
that must be present in the model to complete the UV behaviour. 
In this work, we will show that just the Action (\ref{model}) can
explain the late time cosmology since the first term can account for the
dark energy (DE) content, while the third term is able to explain the dark matter (DM) one.
The first term is just the standard cosmological constant, that we will neglect along
this work. We will focus on the new phenomenology that introduces the third term when it can be
identified with the observed DM. 
\begin{figure}[bt]
\begin{center}
\resizebox{8.0cm}{!} {\includegraphics{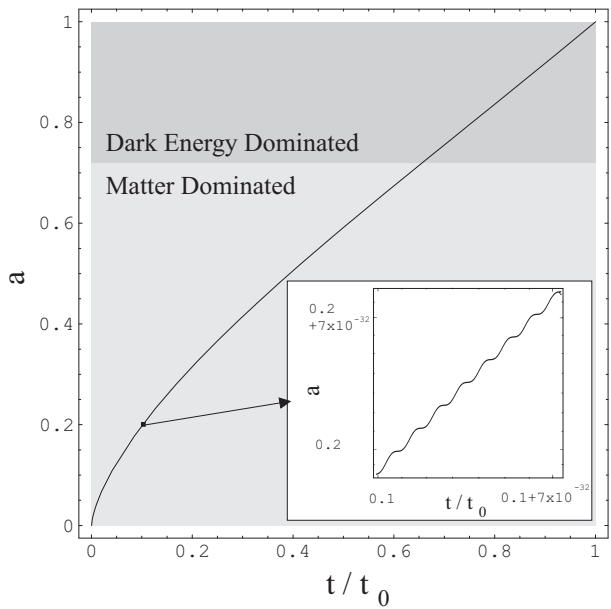}}
\caption{Evolution of the scale factor of the Universe: $a(t)$ as function of time $t$ (time is
normalized to the age of the Universe $t_0\simeq 4.3\times 10^{17}\,s$ \cite{Komatsu:2008hk},
and $a(t_0)=1$). The standard evolution is modulated by
a coherent oscillation. Although this oscillation has a very small amplitude, it has associated a 
high frequency (determined by the mass of the scalar mode, $m_0\simeq1\,\ev$ in the figure) 
and it can constitute the observed amount of DM (we do not take into account the inhomogeneities coming from the
clusterization process).}
\label{Oscp}
\end{center}
\end{figure}
$R^2$-gravity modifies Einstein's Equations (EEs) as \cite{Starobinsky:1980te,Gottlober:1990um}
(following notation from \cite{Cembranos:2005fi}):
\begin{eqnarray}\label{equation1}
&&\left[1-\frac{1}{3\,m_0^2}\,R\right]R_{\mu\nu}-
\frac{1}{2}\left[R-\frac{1}{6\,m_0^2}\,R^2\right]g_{\mu\nu}\nonumber\\
&&
-\,\,\I_{\alpha\beta\mu\nu}
\nabla^\alpha\nabla^\beta\left[\frac{1}{3\,m_0^2}\,R\right]
=\frac{T_{\mu\nu}}{M_{\pla}^2}\,,
\end{eqnarray}
where $\I_{\alpha\beta\mu\nu}\equiv \left(g\dab\gd-g_{\alpha\mu}g_{\beta\nu}\right)$.
The new terms do not modify the standard EEs at low energies except for the mentioned introduction
of a new mode. In fact, if we impose to preserve standard gravity up to nuclear densities or 
Big Bang Nucleosynthesis (BBN) temperatures, the constraints on $m_0$ are just $m_0 \gtrsim 10^{-12}~\ev$.
In this case, the new terms are expected to be negligible at densities lower than
$\rho \sim {T_{\bbn}}^4 \sim (100~\Mev)^4$ or curvatures lower than 
${H_{\bbn}}^2\sim{T_{\bbn}}^4/{M_{\pla}}^2 \sim (10^{-12}~\ev)^2$. In this paper, 
we will discuss in detail the restrictions and possible signatures of the model. 
We will see that the ones already discussed are not dominant.

It is straight forward to check that the metric $g_{\mu\nu}=[1+c_1 \sin(m_0 t)] \eta_{\mu\nu}$ is solution of the 
linearized Eq. (\ref{equation1}), i.e. for $c_1\ll 1$, without any kind of energy source.
In this work we will argue that the energy stored in such oscillations behaves exactly as 
cold DM and can explain the missing matter problem of the Universe (see Fig. \ref{Oscp}). 
We want to emphasize that this new mode of the metric is an independent degree of freedom that 
eventually will cluster and generate a successful structure formation if it is produced in the proper amount.

\section{Interaction with Standard Model particles}

In order to be quantitative, we need to write
the action for the new scalar degree of freedom
of the metric in a canonical way. This work can be done 
directly \cite{Stelle:1977ry} (what it is called 
inside the {\it Jordan frame}) or through a conformal
transformation of the metric~\cite{Kalara:1990ar} 
(what it is known as the {\it Einstein frame}).
As we will work in the limit in which $R\ll m_0^2$, 
in both cases, the metric can be expanded perturbatively as
\begin{eqnarray}
\label{metricperturbation}        
g_{\mu\nu}=\hat{g}_{\mu\nu}+\frac2{M_{\pla}}h_{\mu\nu}
-\sqrt{\frac23}\frac1{M_{\pla}}\,\phi\,\hat{g}_{\mu\nu}\,,
\end{eqnarray}
where $\hat{g}_{\mu\nu}$ is its classical background solution, $h_{\mu\nu}$ 
takes into account the standard two degrees of freedom associated with 
the spin-two (traceless) graviton, and $\phi$ corresponds to the new mode. This scalar
field has associated a canonical kinetic term with the mass $m_0$ as we have already commented. 

We will deduce the couplings of this scalar graviton with the 
SM fields by supposing that gravity is minimally coupled to matter (in the Jordan frame). 
In such a case, $\phi$ is linearly coupled to matter through the trace of the 
standard energy-momentum tensor:
\begin{eqnarray}
\label{coupling}        
{\cal L}_{\phi-T_{\mu\nu}}&=&\frac{1}{M_{\pla}\sqrt{6}}\,\phi\,T^{\mu}_{\mu}\,.
\end{eqnarray}
It implies that the couplings with massive SM particles are given a tree level. 
In particular, the three body interactions are given by:
\begin{eqnarray}
\label{mass}
{\cal L}^{\text{tree-level}}_{\phi-SM} &=& \frac{1}{M_{\pla}\sqrt{6}}\,\phi\,
\left\{2\,m_\Phi^2 \Phi^2-\nabla_\mu \Phi\nabla^\mu \Phi 
\right.
\\
&+& 
\sum_\psi m_\psi\, {\bar \psi} \psi
\left.
 - 2\, m_W^2\, W^+_\mu {W^-}^\mu -\,m_Z^2\, Z_\mu Z^\mu
 \right\}\,
\nonumber
\end{eqnarray}
with the Higgs boson ($\Phi$), (Dirac) fermions ($\psi$), and electroweak gauge bosons, respectively. In contrast with what has been claimed in previous studies,
this field does couple to photons and gluons due to the {\it conformal anomaly} induced at one loop by charged fermions and gauge bosons. We find (following notation from \cite{Goldberger:2007zk}):
\begin{eqnarray}
\label{massless}
{\cal L}^{\text{one-loop}}_{\phi-SM} &=&
\frac{1}{M_{\pla}\sqrt{6}}\,\phi\,
\left\{
{\alpha_{EM} c_{EM} \over  8 \pi}\, F_{\mu\nu} F^{\mu\nu}
\right.
\nonumber
\\
&+& 
\left.
{\alpha_s c_G  \over 8\pi}\, G^a_{\mu\nu} G_a^{\mu\nu}
\right\}\,,
\end{eqnarray}
The particular value of the couplings ($c_{EM}$ and $c_G$) depends on the energy.
We will be particularly interested on the coupling with photons, which leads to potential 
observational decays of $\phi$. We will perform all the calculations restricting ourselves to the 
content of the SM but the exact values of the couplings depend also on 
heavier particles, charged with respect to these gauge interactions, that may extend the SM at higher energies. 

\section{Abundance}

In principle, the above interactions of the scalar mode with the SM could produce a thermal abundance of $\phi$
at a very early stage of the Universe. However, it is expected that higher order corrections to Action (\ref{model})  will be important at this point. In fact, it will typically take place at temperatures $T \gg \Lambda_G\equiv\sqrt{M_{\pla} m_0}$, when we need to know the UV completion of the gravitational theory to study its dynamic. Nevertheless, there is at least another abundance source for this scalar mode that can be computed with Eq. (1). In the same case that other bosonic particles, such as axions \cite{axions}, this field may have associated big abundances through the so called {\it misalignment mechanism}. There is no reason to expect that the initial value of the scalar field ($\phi_1$) should coincide with the minimum of its potential ($\phi=0$) if $H(T)\gg m_0$. Below the temperature $T_1$ for which $3H(T_1)\simeq m_0$, $\phi$ behaves as an standard scalar. It oscillates around the minimum. These oscillations correspond to a zero-momentum condensate, whose initial number density: $n_\phi \sim m_0\phi_1^2/2$
(where $\phi_1=\sqrt{\left\langle \phi(T_1)^2\right\rangle}$ ), will evolve as the typical one associated to standard non-relativistic matter.

Taking into account that the number density of scalar particles scales as the entropy density of radiation
($s=2 \pi^2 g_{s1} T_1^3/45$) in an adiabatic expansion, we can write:
\begin{eqnarray}
\Omega_{\phi}h^2&\simeq& 
\frac{(n_\phi/s)(s_0/\gamma_{s1})}{\rho_{crit}}\,m_0\,,
\end{eqnarray}
where $\rho_{crit}\equiv 1.0540\times10^{4}\,\ev\, \cm^{-3}$ is the critical 
density, $s_0=2970\,\cm^{-3}$ is the present entropy density of the radiation,
and $\gamma_{s1}$ is the factor that this entropy has increased in a comoving volume
since the onset of scalar oscillations. 

If we supposed a radiation dominated universe at $T_1$ ($3H_1=\pi(g_{e\,1}/10)^{1/2}T_1^2/M_{\pla}$), 
we can estimate $T_1$ by solving $m_0=3H_1(T_1)$:
\begin{eqnarray}
\label{T1}
T_1
\simeq 15.5\,\tev
\left[
\frac{m_0}{1\, \ev}
\right]^{\frac12}
\left[
\frac{100}{g_{e\,1}}
\right]^\frac14\,,
\end{eqnarray}
and calculate the abundance as:
\begin{eqnarray}
\Omega_{\phi}h^2&\simeq& 
0.86\,
\left[
\frac{m_0}{1\, \ev}
\right]^{\frac12}
\left[
\frac{\phi_1}{10^{12}\, \gev}
\right]^{2}
\left[
\frac{100\, g_{e\,1}^3}{(\gamma_{s1} g_{s1})^4}
\right]^{\frac14},\;\;\;\;\;\;
\label{abundance}
\end{eqnarray}
where $g_{e\,1}$ ($g_{s1}$) are the effective energy (entropy)
number of relativistic degrees of freedom at $T_1$.
We see that initial values for the scalar field of order of 
$\phi_1\sim 10^{12}\;\gev$ can lead to the non-baryonic DM (NBDM) abundance depending 
on the rest of parameters and the early physics of the Universe (see Fig. \ref{gravDM}).
We can check that this result is consistent with a perturbative treatment of the 
background metric $||\Delta g_{\mu\nu}/\hat{g}_{\mu\nu}||\lesssim 10^{-6}$ for the entire computation.

\section{Signatures and constraints}

On the other hand, Eqs. (\ref{coupling},\ref{mass}) imply that the new scalar graviton mediates a standard 
attractive Yukawa force between two non-relativistic particles of masses $M_a$ and $M_b$:
\begin{eqnarray}
\label{Yukawa}        
V_{ab}&=&-\alpha\frac{1}{8 \pi M_{\pla}^2}\frac{M_a M_b}{r} e^{- m_0\,r}\,,
\end{eqnarray}
with $\alpha=1/3$ \cite{Stelle:1977ry}. The non-observation of such a force by torsion-balance
experiments requires \cite{Kapner:2006si,Adelberger:2006dh}:
\begin{eqnarray}
\label{yukawa}        
m_0\geq 2.7 \times 10^{-3} \ev \;\;\;\;\;\;\;\text{at 95 \% c.l.}
\end{eqnarray}
This is the most constraining lower bound on the mass of the scalar mode
and it is independent of its misalignment or any other supposition about
its abundance.

On the contrary, depending on its abundance, $m_0$ is constrained from above.
It is particularly interesting the decay in $e^+e^-$ since it is the most constraining
if $\phi$ constitutes the total NBDM. From (\ref{mass}), it is possible
to calculate the $\phi$ decay rate into a generic pair fermion anti-fermion.
We find:
\begin{eqnarray}
\label{fermion}        
\Gamma_{\phi\rightarrow \psi \bar{\psi}}=\frac{m_\psi^2 m_0\,N_c}{48 \pi M_{\pla}^2\,\mu}\left(1-\frac{4 m_\psi^2}{m_0^2}\right)^{3/2}\,,
\end{eqnarray}
where $N_c$ is the number of colors, and $\mu=1$ or $\mu=2$ depending if the fermion is Dirac or Majorana type. 
In particular, for the electron-positron decay ($N_c=1$, $\mu=1$ and defining $r_e=m_0/(2 m_{e})$):
\begin{eqnarray}
\label{positron}        
\Gamma_{\phi\rightarrow e^+e^-}\simeq 
\left[
2.14 \times 10^{24} s\; \frac{r_e^2}{\left(r_e^2-1\right)^{3/2}}
\right]^{-1}
\,.
\end{eqnarray}
Restrictions are set by the observations of
the SPI spectrometer on the INTEGRAL (International Gamma-ray Astrophysics Laboratory) 
satellite, which has measured a 511 keV line emission of $1.05 \pm 0.06 \times 10^{-3}$ photons 
cm$^{-2}$ s$^{-1}$ from the Galactic center (GC)~\cite{Knodlseder:2005yq}, confirming previous measurements. 
This 511 keV line flux is fully consistent with an $e^+ e^-$ annihilation spectrum although the 
source of the positrons is unknown.

\begin{figure}[bt]
\begin{center}
\resizebox{8.0cm}{!} {\includegraphics{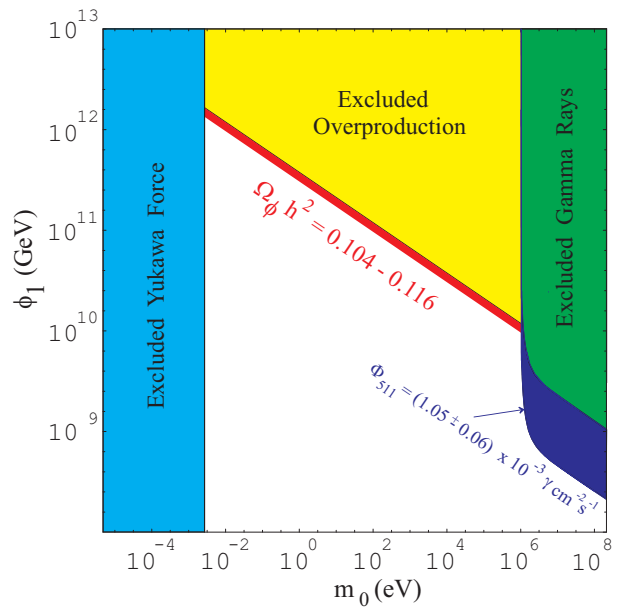}}
\caption{Parameter space of the model: $m_0$ is the mass of the new scalar mode and 
$\phi_1$ is its misalignment when $3H\sim m_0$ (we assume $g_{e\,1}=g_{s1}\simeq 106.75$, and $\gamma_{s1}\simeq 1$). 
The left side is excluded by modifications of Newton's law. The right one is excluded by cosmic-ray observations. 
In the limit of this region, $R^2$-gravity can account for the positron production in order to explain the 511 keV 
line coming from the GC confirmed by INTEGRAL \cite{Knodlseder:2005yq} (up to $m_0 \sim 10\; \Mev$). The upper area 
is ruled out by DM overproduction. The diagonal line corresponds to the NBDM abundance fitted with WMAP data 
\cite{Komatsu:2008hk}.
}
\label{gravDM}
\end{center}
\end{figure}

If $m_0 \geq 1.2\; \Mev$, the scalar field cannot constitute the total local DM since
we should observe a bigger excess of the 511 line coming from the GC. On the other hand, 
decaying DM (DDM) has been already proposed in different works as a possible source of the 
inferred positrons if its mass is lighter than $M_{\text{DDM}}\lesssim 10\;\Mev$ \cite{Beacom:2005qv} 
and its decay rate in $e^+e^-$ verifies 
\cite{Picciotto:2004rp, Hooper:2004qf, Kasuya:2006kj, Pospelov:2007xh, Cembranos:2008bw}: 
\begin{eqnarray}
\label{511}        
\frac{\Omega_{\text{DDM}}h^2\; \Gamma_{\text{DDM}}}{M_{\text{DDM}}} \simeq 
\left[(0.2 - 4)\times 10^{27}\; \s\; \Mev\right]^{-1}
\,.
\end{eqnarray}
The most important uncertainty for this interval comes from the dark halo profile, although
a cuspy density is definitely needed (with a inner slope $\gamma\gtrsim 1.5$ \cite{Cembranos:2008bw}).
If $m_0$ is tunned to $2\,m_e$ with an accuracy of $5-10\; \%$, the line could be explained by $R^2$-gravity.
The same gravitational DM can explain the 511 line with a less tunned mass (up to $m_0 \sim 10\; \Mev$)
if the misalignment is $\phi_1 \sim 10^9\; \gev$, i.e. with a lower abundance (See Fig. \ref{gravDM}).
If $m_0 \gtrsim 10\; \Mev$, the gamma ray spectrum originated by inflight annihilation of the positrons with interstellar electrons is even more constraining than the 511 keV photons \cite{Beacom:2005qv}.
 
On the contrary, if $m_0 < 2\, m_e$, the only decay channel that may be observable is the decay in two 
photons. We find:
\begin{eqnarray}
\label{foton}        
\Gamma_{\phi\rightarrow \gamma\gamma}=\frac{\alpha_{EM}^2 m_0^3}{1536 \pi^3 M_{\pla}^2}\left|c_{EM}\right|^{2}\,.
\end{eqnarray}
By taking into account all SM charged particles and assuming $\phi$ to be much lighter
than all of them: $c_{EM}=11/3$, and
\begin{eqnarray}
\label{foton2}        
\Gamma_{\phi\rightarrow \gamma\gamma}
\simeq \left[2.5 \times 10^{29} s\left[\frac{1\, \Mev}{m_0}\right]^3\right]^{-1}
.\;\;\;\;
\end{eqnarray}
As it has been discussed in detail in \cite{Cembranos:2007fj,Cembranos:2008bw}, 
if $m_0\lesssim 1\; \Mev$, it is difficult to detect these gravitational decays in  the isotropic 
diffuse photon background (iDPB). The gamma-ray spectrum at high Galactic latitudes 
can have contributions from Galactic and extragalactic sources, but it seems well
fitted at $E_\gamma\lesssim 1\; \Mev$ by assuming Active Galactic Nuclei (AGN) as main sources. 
The spectrum observed by COMPTEL (-the Compton Imaging Telescope- over the energy ranges 
$0.8-30~\Mev$~\cite{Weidenspointner}), SMM (-the Solar Maximum Mission- for $0.3-7~\Mev$~\cite{WatanabeSMM})
and INTEGRAL ( $5-100~\kev$~\cite{Churazov:2006bk}), fall like a power law, 
with $dN/dE \sim E^{-2.4}$~\cite{Weidenspointner}, and it will dominate 
any possible signal from $R^2$-gravity if $m_0\lesssim 1\;\Mev$.

However, a most promising analysis is associated with the search of gamma-ray lines 
at $E_\gamma= m_0/2$ from localized sources, as the GC. The iDPB is 
continuum since it suffers the cosmological redshift. However, the mono-energetic photons originated
by local sources may give a clear signal of $R^2$-gravity. INTEGRAL has performed 
a search for gamma-ray lines originated within $13^\circ$ 
from the GC over the energy ranges 
$0.08-8~\Mev$. It has not observed any line below 511 keV up to upper flux limits of 
$10^{-5}$-$10^{-2}$ cm$^{-2}$ s$^{-1}$, depending on line width, energy, and exposure \cite{Teegarden:2006ni}. 
Unfortunately, these flux limits are, at least, one order of magnitude over the expected 
fluxes from $\phi$ decays with $m_0 \lesssim 1\; \Mev$, even for cuspy halos.
The photon flux originated by $R^2$-gravity depends on $m_0$ as 
$\Phi_{E_{\gamma}=m_0/2}\,\propto \, m_0^2$. This strong dependence implies
that only the heavier allowed region could be detected with reasonable improvements of present experiments \cite{Cembranos:2007fj}.

\section{Conclusions}

In conclusion, we have studied the possibility that the DM origin resides in UV modifications of gravity. 
Although our results may seem particular of $R^2$-gravity, the low energy phenomenology 
of the studied scalar mode is ubiquitous in high energy corrections of the EHA coming from string theory, 
supersymmetry or extra dimensions (reproduced in form of dilatons, radions, graviscalars or other moduli fields). 
The instability of the deduced DM predicts a deviation from EEs at an energy scale: 
$1\,\tev \lesssim \Lambda_G\lesssim 10^5\,\tev$. Consequences for hierarchy interpretations, 
baryogenesis or inflation deserve further investigations.

\section*{Acknowledgments}
\vspace{.2 cm}
I am grateful to K. Olive, M. Peloso, and M. Voloshin for useful discussions. This work is supported 
in part by DOE grant DOE/DE-FG02-94ER40823, FPA 2005-02327 project (DGICYT, Spain), and CAM/UCM 910309 project.

\end{document}